\documentclass{article}

\usepackage[numbers]{natbib}
\usepackage[preprint]{neurips_2022}
\usepackage{amsmath}
\usepackage{amssymb}
\usepackage{mathtools}
\usepackage{amsthm}
\DeclareSymbolFont{extraup}{U}{zavm}{m}{n}
\DeclareMathSymbol{\varheart}{\mathalpha}{extraup}{86}
\DeclareMathSymbol{\vardiamond}{\mathalpha}{extraup}{87}

\usepackage[utf8]{inputenc} % allow utf-8 input
\usepackage[T1]{fontenc}    % use 8-bit T1 fonts
\usepackage{hyperref}       % hyperlinks
\usepackage{url}            % simple URL typesetting
\usepackage{booktabs}       % professional-quality tables
\usepackage{amsfonts}       % blackboard math symbols
\usepackage{nicefrac}       % compact symbols for 1/2, etc.
\usepackage{microtype}      % microtypography
\usepackage[dvipsnames]{xcolor}
\usepackage{amsmath}
\usepackage{graphicx}
\usepackage{graphics}
\usepackage{subcaption}
\usepackage{accents}
\usepackage{enumitem}
\usepackage{tabu}
\usepackage{wrapfig}
\usepackage{graphicx}
\usepackage{multirow}
\usepackage{multicol}
\usepackage{float}
\usepackage{color}
\usepackage{caption}
\usepackage{bbding}
\usepackage{hanging}

\usepackage{tablefootnote}
\usepackage{diagbox}

\def\myname{NaturalSpeech 3}

\title{\textit{NaturalSpeech 3}: Zero-Shot Speech Synthesis\\with Factorized Codec and Diffusion Models}

\author{
Zeqian Ju$^1{}^2$\thanks{The first four authors contributed equally to this work, and their names are listed in random order. Corresponding author: Xu Tan, \texttt{xuta@microsoft.com}}, ~Yuancheng Wang$^3$\footnotemark[1], ~Kai Shen$^4{}^1$\footnotemark[1], ~Xu Tan$^1$\footnotemark[1], ~Detai Xin$^1{}^5$, Dongchao Yang$^1$, \\
\textbf{Yanqing Liu$^1$, Yichong Leng$^1$, Kaitao Song$^1$, Siliang Tang$^4$, Zhizheng Wu$^3$, Tao Qin$^1$,} \\
\textbf{Xiang-Yang Li$^2$, Wei Ye$^6$, Shikun Zhang$^6$, Jiang Bian$^1$, Lei He$^1$, Jinyu Li$^1$, Sheng Zhao$^1$} \\ 
$^1$Microsoft Research \& Microsoft Azure \\
$^2$University of Science and Technology of China $^3$The Chinese University of Hong Kong, Shenzhen \\
$^4$Zhejiang University, $^5$The University of Tokyo, $^6$Peking University \\
\url{https://aka.ms/speechresearch} 
}
\begin{document}

\maketitle

\vspace{-0.3cm}

\begin{abstract}
While recent large-scale text-to-speech (TTS) models have achieved significant progress, they still fall short in speech quality, similarity, and prosody. Considering speech intricately encompasses various attributes (e.g., content, prosody, timbre, and acoustic details) that pose significant challenges for generation, a natural idea is to factorize speech into individual subspaces representing different attributes and generate them individually. Motivated by it, we propose \textit{\myname{}}, a TTS system with novel factorized diffusion models to generate natural speech in a zero-shot way. Specifically, 1) we design a neural codec with factorized vector quantization (FVQ) to disentangle speech waveform into subspaces of content, prosody, timbre, and acoustic details; 2) we propose a factorized diffusion model to generate attributes in each subspace following its corresponding prompt. With this factorization design, \myname{} can effectively and efficiently model intricate speech with disentangled subspaces in a divide-and-conquer way. Experiments show that \myname{} outperforms the state-of-the-art TTS systems on quality, similarity, prosody, and intelligibility, and achieves on-par quality with human recordings. Furthermore, we achieve better performance by scaling to 1B parameters and 200K hours of training data.

\end{abstract}

\begin{figure}[H]  
  \centering  
  \resizebox{0.87\textwidth}{!}{%
  \begin{tabular}{c c}  
    \multirow{2}{*}[0em]{  
      \begin{subfigure}[t]{0.50\linewidth}  
        \caption{}  
        \includegraphics[page=1,width=\linewidth,trim=0.0cm 3.49cm 10.55cm 0.0cm,clip=true]{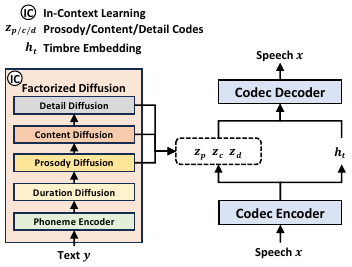}  
      \end{subfigure}  
    } &  
    \begin{subfigure}[t]{0.40\linewidth}  
      \caption{}  
      \includegraphics[page=1,width=\linewidth]{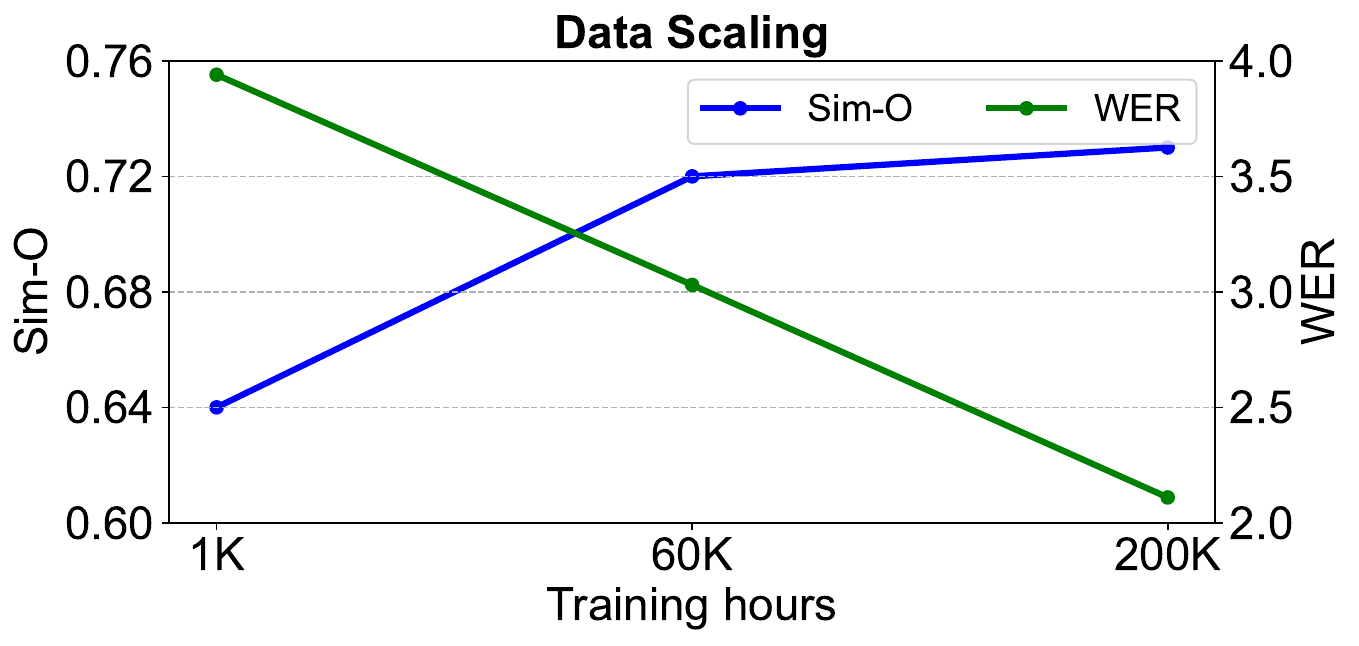}  
      \includegraphics[page=1,width=\linewidth]{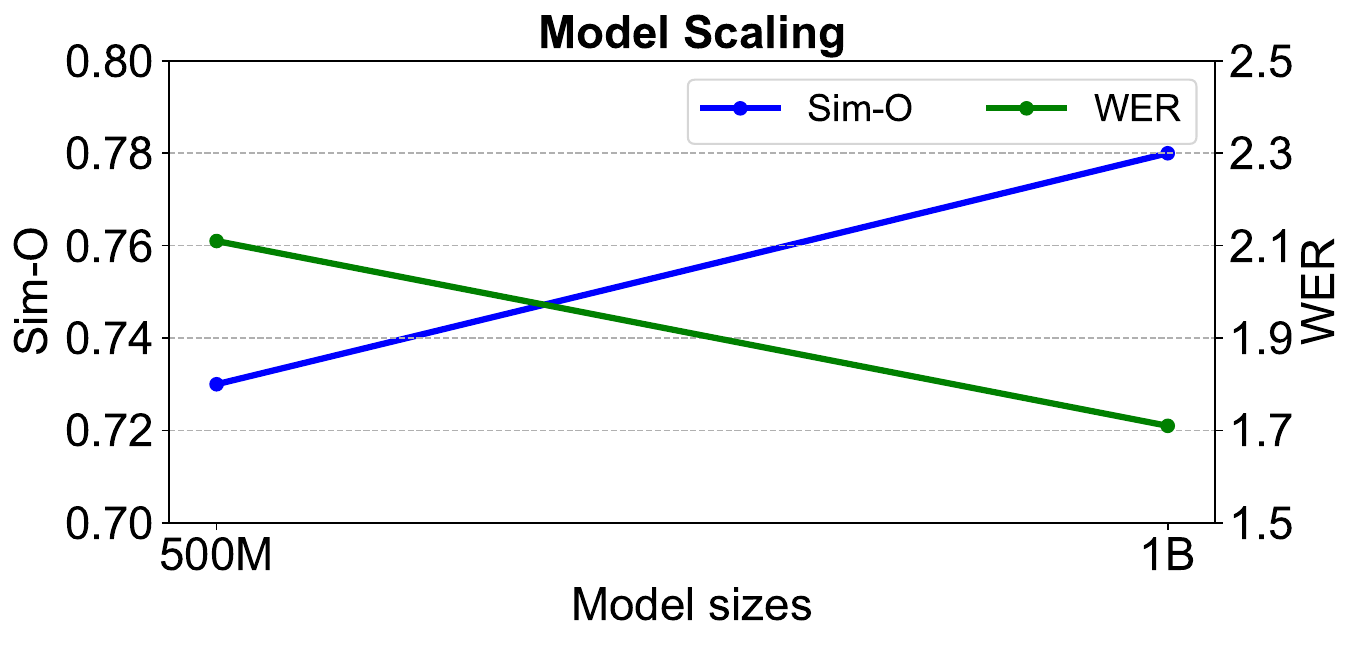}  
    \end{subfigure} \\ 
  \end{tabular}  
  }
  \caption{(a) Overview of the system, with a neural speech codec for speech attribute factorization and a factorized diffusion model. (b) Data and model scaling of the system.}  
  \label{fig_system_overview}  
\end{figure} 

\section{Introduction}

In recent years, significant advancements have been achieved in text-to-speech (TTS) synthesis. Traditional TTS systems~\cite{wang2017tacotron, shen2018natural, ren2019fastspeech, tan2024naturalspeech} are typically trained on limited datasets recorded in studios, and thus fail to support high-quality zero-shot speech synthesis. Recent works~\cite{shen2023naturalspeech, wang2023neural, jiang2023mega} have made considerable progress for zero-shot TTS by largely scaling up both the corpus and the model sizes. However, the synthesis results of these large-scale TTS systems are not satisfactory in terms of voice quality, similarity, and prosody.

The challenges of inferior results stem from the intricate information embedded in speech, since speech encompasses numerous attributes, such as content, prosody, timbre, and acoustic detail. Previous works using raw waveform~\cite{kim2021conditional, lim2022jets} and mel-spectrogram~\cite{wang2017tacotron, shen2018natural, popov2021grad, jiang2023mega, le2023voicebox} as data representations suffer from these intricate complexities during speech generation. A natural idea is to factorize speech into disentangled subspaces representing different attributes and generate them individually. However, achieving this kind of disentangled factorization is non-trivial. Previous works~\cite{borsos2022audiolm, borsos2023soundstorm, wang2023neural} encode speech into multi-level discrete tokens using a neural audio codec~\cite{zeghidour2021soundstream, defossez2022high} based on residual vector quantization (RVQ). Although this approach decomposes speech into different hierarchical representations, it does not effectively disentangle the information of different attributes of speech across different RVQ levels and still suffers from modeling complex coupled information.

To effectively generate speech with better quality, similarity and prosody, we propose a TTS system with novel factorized diffusion models to generate natural speech in a zero-shot way. Specifically, 1) we introduce a novel neural speech codec with factorized vector quantization (FVQ), named FACodec,  to decompose speech waveform into distinct subspaces of content, prosody, timbre, and acoustic details and reconstruct speech waveform with these disentangled representations, leveraging information bottleneck~\cite{qian2020unsupervised, qian2019autovc}, various supervised losses, and adversarial training~\cite{kong2020hifi} to enhance disentanglement; 2) we propose a factorized diffusion model, which generates the factorized speech representations of duration, content, prosody, and acoustic detail, based on their corresponding prompts. This design allows us to use different prompts to control different attributes. The overview of our method, referred to \myname{}, is shown in Figure~\ref{fig_system_overview}.

We decompose complex speech into subspaces representing different attributes, thus simplifying the modeling of speech representation. This approach offers several advantages: 1) our factorized diffusion model is able to learn these disentangled representations efficiently, resulting in higher quality speech generation; 2) by disentangling timbre information in our FACodec, we enable our factorized diffusion model to avoid directly modeling timbre. This reduces learning complexity and leads to improved zero-shot speech synthesis; 3) we can use different prompts to control different attributes, enhancing the controllability of \myname{}.

Benefiting from these designs, \myname{} has achieved significant improvements in speech quality, similarity, prosody, and intelligibility. Specifically, 1) it achieves comparable or better speech quality than the ground-truth speech on the LibriSpeech test set in terms of CMOS; 2) it achieves a new SOTA on the similarity between the synthesized speech and the prompt speech (0.64 $\rightarrow$ 0.67 on Sim-O, 3.69 $\rightarrow$ 4.01 on SMOS); 3) it shows a significant improvement in prosody compared to other TTS systems with $-$0.16 average MCD (lower is better), $+$0.21 SMOS; 4) it achieves a SOTA on intelligibility (1.94 $\rightarrow$ 1.81 on WER); 5) it achieves human-level naturalness on multi-speaker datasets (e.g., LibriSpeech), another breakthrough after NaturalSpeech\footnote{While NaturalSpeech 1~\cite{tan2024naturalspeech} achieved human-level quality on the single-speaker LJSpeech dataset, NaturalSpeech 3 achieved human-level quality on the diverse multi-speaker LibriSpeech dataset for the first time.}. Furthermore, we demonstrate the scalability of \myname{} by scaling it to 1B parameters and 200K hours of training data.
Audio samples can be found in \url{https://speechresearch.github.io/naturalspeech3}.

\section{Background}
In this section, we discuss the recent progress in TTS including: 1) zero-shot TTS; 2) speech representations in TTS; 3) generation methods in TTS; 4) speech attribute disentanglement.

\textbf{Zero-shot TTS.} Zero-shot TTS aims to synthesize speech for unseen speakers with speech prompts. We can systematically categorize these systems into four groups based on data representation and modelling methods: 1) Discrete Tokens + Autoregressive~\cite{wang2023neural, kharitonov2023speak, huang2023make}; 2) Discrete Tokens + Non-autoregressive~\cite{borsos2023soundstorm, yang2023instructtts, du2023unicats}; 3) Continuous Vectors + Autoregressive~\cite{nachmani2023lms};  4) Continuous Vectors + Non-autoregressive~\cite{shen2023naturalspeech, le2023voicebox, li2023styletts, lee2023hierspeech++}. Discrete tokens are typically derived from neural codec, while continuous vectors are generally obtained from mel-spectrogram or latents from audio autoencoder or codec. In addition to the aforementioned perspectives, we disentangle speech waveforms into subspaces based on attribute disentanglement and propose a factorized diffusion model to generate attributes within each subspace, motivated by the principle of divide-and-conquer. Meanwhile, we can reuse previous methods, employing discrete tokens along with autoregressive models.

\textbf{Speech Representations in TTS.} Traditional works propose using prior-based speech representation such as raw waveform \cite{oord2016wavenet, oord2018parallel, sotelo2017char2wav} or mel-spectrogram \cite{ping2018deep, li2019neural, ren2019fastspeech, kim2020glow}. Recently, large-scale TTS systems \cite{wang2023neural, borsos2023soundstorm, shen2023naturalspeech} leverage data-driven representation, i.e., either discrete tokens or continuous vectors form an auto-encoder \cite{zeghidour2021soundstream, defossez2022high, kumar2023high}. However, these methods ignore that speech contains various complex attributes and encounter intricate complexities during speech generation. In this paper, we factorize speech into individual subspaces representing different attributes which can be effectively and efficiently modeled.

\textbf{Generation Methods in TTS.} Previous works have demonstrated that NAR-based models~\cite{ren2019fastspeech, elias2020parallel, liu2022diffsinger, jiang2023mega, shen2023naturalspeech, le2023voicebox} enjoy better robustness and generation speed than AR-based models, because they explicitly model the duration and predict all features simultaneously. Instead, AR-based models~\cite{shen2018natural, li2019neural, wang2023neural, nachmani2023lms, yang2023uniaudio} have better diversity, prosody, expressiveness, and flexibility than NAR-based models, due to their implicitly duration modeling and token sampling strategy. In this study, we adopt the NAR modeling approach and propose a factorized diffusion model to support our disentangled speech representations and also extend it to AR modeling approaches. This allows \myname{} to achieve better expressiveness while maintaining stability and generation speed.

\textbf{Speech Attribute Disentanglement.} Prior works~\cite{choi2021neural, choi2022nansy++, polyak2021speech} utilize disentangled representation for speech generation, such as speech content from self-supervised pre-trained models~\cite{chung2021w2v, baevski2020wav2vec, schneider2019wav2vec}, fundamental frequency, and timbre, but  speech quality is not satisfying. Recently, some works explore attribute disentanglement in neural speech codec. SpeechTokenizer~\cite{zhang2023speechtokenizer} uses HuBERT~\cite{hsu2021hubert} for semantic distillation, aiming to render the first-layer RVQ representation as semantic information. Disen-TF-Codec~\cite{jiang2023disentangled} proposes the disentanglement with content and timbre representation, and applies them for zero-shot voice conversion. In this paper, we achieve better disentanglement with more speech attributes including content, prosody, acoustic details and timbre while ensuring high-quality reconstruction. We validate such disentanglement can bring about significant improvements in  zero-shot TTS task.

\section{\myname{}}
\label{sec: method}

\subsection{Overall Architecture}
\label{sec: method-overview}

In this section, we present \myname{}, a cutting-edge system for natural and zero-shot text-to-speech synthesis with better speech quality, similarity and controllability.
As shown in Figure \ref{fig_system_overview}, \myname{} consists of 1) a neural speech codec (i.e., FACodec) for attribute disentanglement; 2) a factorized diffusion model which generates factorized speech attributes.  
Since the speech waveform is complex and intricately encompasses various attributes, we factorize speech into five attributes including: duration, prosody, content, acoustic details, and timbre. Specifically, although the duration can be regarded as an aspect of prosody, we choose to model it explicitly due to our non-autoregressive speech generation design. We use our internal alignment tool to alignment speech and phoneme and obtain phoneme-level duration. For other attributes, we implicitly utilize the factorized neural speech codec to learn disentangled speech attribute subspaces (i.e., content, prosody, acoustic details, and timbre).
Then, we use the factorized diffusion model to generate each speech attribute representation. 
Finally, we employ the codec decoder to reconstruct the waveform with the generated speech attributes. We introduce the FACodec in Section \ref{sec: method-codec} and the factorized diffusion model in Section \ref{sec: method-gen}.

\subsection{FACodec for Attribute Factorization}
\label{sec: method-codec}

\subsubsection{FACodec Model Overview}
\label{sec: codec-overview}

We propose a factorized neural speech codec (i.e., FACodec\footnote{
We release the code and pre-trained checkpoint of FACodec at \url{https://huggingface.co/spaces/amphion/naturalspeech3_facodec}.}) to convert complex speech waveform into disentangled subspaces representing speech attributes of content, prosody, timbre, and acoustic details and reconstruct high-quality speech waveform from these.

\begin{figure*}[t]
  \centering
  % \vspace{-10px}
  \includegraphics[page=1,width=0.9\columnwidth,trim=0cm 3.6cm 3.5cm 0.0cm,clip=true]{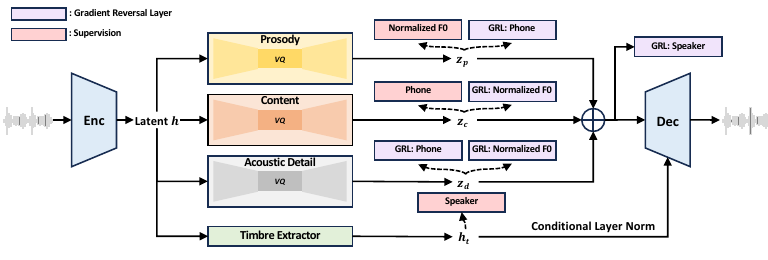}
  \caption{The framework of the FACodec for attribute factorization.}
  \label{fig_codec}
\end{figure*}

As shown in Figure~\ref{fig_codec}, our FACodec consists of a speech encoder, a timbre extractor, three factorized vector quantizers (FVQ) for content, prosody, acoustic detail, and a speech decoder. Given a speech $x$,
1) following \cite{zeghidour2021soundstream,shen2023naturalspeech}, we adopt several convolutional blocks for the speech encoder with a downsample rate of 200 for 16KHz speech data (i.e., each frame corresponding to a 12.5ms speech segment) to obtain pre-quantization latent $h$;
2) the timbre extractor is a Transformer encoder which converts the output of the speech encoder $h$ into a global vector $h_t$ representing the timbre attributes;
3) for other attribute $i$ ($i = {p, c, d}$ for prosody, content, and acoustic detail, respectively), we use a factorized vector quantizer ($\text{FVQ}_i$) to capture fine-grained speech attribute representation and obtain corresponding discrete tokens;
4) the speech decoder mirrors the structure of speech encoder but with much larger parameter amount to ensure high-quality speech reconstruction. We first add the representation of prosody, content, and acoustic details together and then fuse the timbre information by conditional layer normalization~\cite{chen2021adaspeech} to obtain the input $z$ for the speech decoder. We discuss how to achieve better speech attribute disentanglement in the next section.

\subsubsection{Attribute Disentanglement}
\label{sec: codec-design-disentanglement-technique}

Directly factorizing speech into different subspaces does not guarantee the disentanglement of speech. In this section, we introduce some techniques to achieve better speech attribute disentanglement: 1) information bottleneck, 2) supervision, 3) gradient reverse, and 4) detail dropout. Please refer to Appendix~\ref{appendix: codec-implement} for more training details.

\textbf{Information Bottleneck.}
Inspired by \cite{qian2020unsupervised, qian2019autovc}, to force the model to remove unnecessary information (such as prosody in content subspace), we construct the information bottleneck in prosody, content, and acoustic details FVQ by projecting the encoder output into a low-dimensional space (i.e., 8-dimension) and subsequently quantize within this low-dimensional space. This technique ensures that each code embedding contains less information, facilitating information disentanglement~\cite{kumar2023high, yu2021vector}. After quantization, we will project the quantized vector back to original dimension.

\textbf{Supervision.} To achieve high-quality speech disentanglement, we introduce supervision as auxiliary task for each attribute. For prosody, since pitch is an important part of prosody~\cite{choi2022nansy++}, 
we take the post-quantization latent $z_p$ to predict pitch information. We extract the F0 for each frame and use normalized F0 (z-score) as the target.
% $\text{F0} - \text{mean}(\text{F0})/\text{std}(\text{F0})$
For content, we directly use the phoneme labels as the target (we use our internal alignment tool to get the frame-level phoneme labels).
For timbre, we apply speaker classification on $h_t$ by predicting the speaker ID.

\textbf{Gradient Reversal.} 
Avoiding the information leak (such as the prosody leak in content) can enhance disentanglement. Inspired by \cite{yang2022speech}, we adopt adversarial classifier with the gradient reversal layer (GRL)~\cite{ganin2015unsupervised} to eliminate undesired information in latent space. Specifically, for prosody, we apply phoneme-GRL (i.e., GRL layer by predicting phoneme labels) to eliminate content information; for content, since the pitch is an important aspect of prosody, we apply F0-GRL to reduce the prosody information for simplicity; for acoustic details, we apply both phoneme-GRL and F0-GRL to eliminate both content and prosody information. In addition, we apply speaker-GRL on the sum of $z_p, z_c, z_d$ to eliminate timbre.

\begin{figure*}[t]
  \centering
  \includegraphics[width=0.98\linewidth]{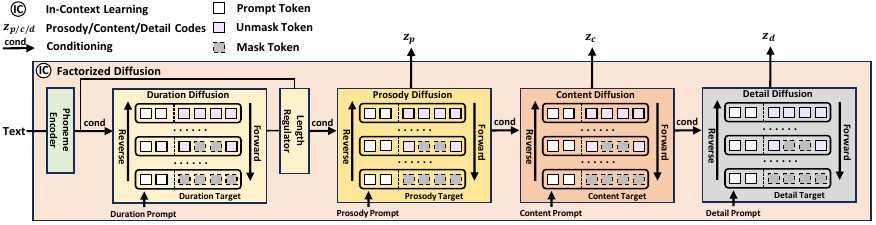 }
  \caption{The framework of factorized diffusion model, which consists of 1) phoneme encoder, 2) duration diffusion and length regulator, 3) prosody diffusion, 4) content diffusion, 5) detail (acoustic detail) diffusion. Note that modules 2-5 share the same diffusion formulation.}
  \label{pic:fdm_framework}
\end{figure*}

\textbf{Detail Dropout.} 
We have the following considerations: 1) empirically, we find that the codec tends to preserve undesired information (e.g., content, prosody) in acoustic details subspace since there is no supervision; 2) intuitively, without acoustic details, the decoder should reconstruct speech only with prosody, content and timbre, although in low-quality. Motivated by them, we design the detail dropout by randomly masking out $z_d$ during the training process with probability $p$. With detail dropout, we achieve the trade-off of disentanglement and reconstruction quality: 1) the codec can fully utilize the prosody, content and timbre information to reconstruct the speech to ensure the decouple ability, although in low-quality; 2) we can obtain high-quality speech when the acoustic details are given.

\subsection{Factorized Diffusion Model }
\label{sec: method-gen}

\subsubsection{Model Overview}
\label{sec:method-speech-generation-overview}

We generate speech with discrete diffusion for better generation quality. We have the following considerations: 1) we factorize speech into the following attributes: duration, prosody, content, and acoustic details, and generate them in sequential with specific conditions. Firstly, as we mentioned in Section \ref{sec: method-overview}, due to our non-autoregressive generation design, we first generate duration. Secondly, intuitively, the acoustic details should be generated at last;
2) following the speech factorization design, we only provide the generative model with the corresponding attribute prompt and apply discrete diffusion in its subspace;
3) to facilitate in-context learning in diffusion model, we utilize the codec to factorize speech prompt into attribute prompts (i.e., content, prosody and acoustic details prompt) and generate the target speech attribute with partial noising mechanism following \cite{gong2022diffuseq,borsos2023soundstorm}. For example, for prosody generation, we directly concatenate prosody prompt (without noise) and target sequence (with noise) and gradually remove noise from target sequence with prosody prompt.

With these thoughts, as shown in Figure \ref{pic:fdm_framework}, we present our factorized diffusion model, which consists of a phoneme encoder and speech attribute (i.e., duration, prosody, content, and acoustic details) diffusion modules with the same discrete diffusion formulation: 1) we generate the speech duration by applying duration diffusion with duration prompt and phoneme-level textural condition encoded by phoneme encoder. Then we apply the length regulator to obtain frame-level phoneme condition $c_{ph}$; 2) we generate prosody $z_p$ with prosody prompt and phoneme condition $c_{ph}$; 3) we generate content prosody $z_c$ with content prompt and use generated prosody $z_p$ and phoneme $c_{ph}$ as conditions; 4) we generate acoustic details $z_d$ with acoustic details prompt and use generated prosody, content and phoneme $z_p, z_c, c_{ph}$ as conditions.
Specifically,
we do not explicitly generate the timbre attribute. Due to the factorization design in our FACodec, we can obtain timbre from the prompt directly and do not need to generate it.
Finally, we synthesize the target speech by combining attributes $z_p, z_c, z_d$ and $h_t$ and decoding it with codec decoder. We discuss the diffusion formulation in Section \ref{sec:method-diffusion-formulation}.

\subsubsection{Diffusion Formulation}
\label{sec:method-diffusion-formulation}

\textbf{Forward Process.} Denote $\mathbf{X}=[x_i]_{i=1}^{N}$ the target discrete token sequence, where $N$ is the sequence length, $\mathbf{X}^{p}$ is the prompt discrete token sequence, and $\mathbf{C}$ is the condition. The forward process at time $t$ is defined as masking a subset of tokens in $\mathbf{X}$ with the corresponding binary mask $\mathbf{M}_{t}=[m_{t,i}]_{i=1}^{N}$, formulated as $\mathbf{X}_{t} = \mathbf{X} \odot \mathbf{M}_{t}$, by replacing $x_i$ with [MASK] token if $m_{t,i}=1$, and otherwise leaving $x_i$ unmasked if $m_{t,i}=0$. $m_{t,i} \stackrel{iid}{\sim} \text{Bernoulli}(\sigma(t))$ and $\sigma(t) \in (0,1]$ is a monotonically increasing function. In this paper, $\sigma(t)=\sin(\frac{\pi t}{2T}), t \in (0,T]$. Specially, we denote $\mathbf{X}_{0}=\mathbf{X}$ for the original token sequence and $\mathbf{X}_{T}$ for the fully masked sequence.

\textbf{Reverse Process.} The reverse process gradually restores $\mathbf{X}_{0}$ by sampling from reverse distribution $q(\mathbf{X}_{t-\Delta t}|\mathbf{X}_{0}, \mathbf{X}_{t})$, starting from full masked sequence $\mathbf{X}_{T}$. Since $\mathbf{X}_{0}$ is unavailable in inference, we use the diffusion model $p_{\theta}$, parameterized by $\theta$, to predict the masked tokens conditioned on $\mathbf{X}^p$ and $\mathbf{C}$, denoted as $p_{\theta}(\mathbf{X}_{0}|\mathbf{X}_{t}, \mathbf{X}^{p}, \mathbf{C})$. The parameters $\theta$ are optimized to minimize the negative log-likelihood of the masked tokens:
\begin{equation*}                    
    \begin{aligned}
        \mathcal{L}_{\text{mask}} &= \mathop{\mathbb{E}}\limits_{\mathbf{X} \in \mathcal{D}, t \in \left[0, T\right]}  -\sum_{i=1}^{N} m_{t,i} \cdot \log(p_{\theta}(x_i|\mathbf{X}_{t}, \mathbf{X}^{p}, \mathbf{C})).
    \end{aligned}
\end{equation*}
Then we can get the reverse transition distribution:
\begin{footnotesize}
    \begin{equation*}
        \begin{aligned}
            p(\mathbf{X}_{t-\Delta t}|\mathbf{X}_{t}, \mathbf{X}^{p}, \mathbf{C})= \mathop{\mathbb{E}}\limits_{\mathbf{\hat X}_{0} \sim p_{\theta} (\mathbf{X}_{0}|\mathbf{X}_{t}, \mathbf{X}^{p}, \mathbf{C})} q(\mathbf{X}_{t-\Delta t}|\mathbf{\hat X}_{0}, \mathbf{X}_{t}).
        \end{aligned}
    \end{equation*}
\end{footnotesize}

\textbf{Inference.} During inference, we progressively replace masked tokens, starting from the fully masked sequence $\mathbf{X}_{T}$, by iteratively sampling from $p(\mathbf{X}_{t-\Delta t}|\mathbf{X}_{t}, \mathbf{X}^{p}, \mathbf{C})$. Inspire by \cite{chung2022diffusion, gu2022vector, lezama2022improved}, we first sample $\mathbf{\hat X}_{0}$ from $p_{\theta}(\mathbf{X}_{0}|\mathbf{X}_{t}, \mathbf{X}^{p}, \mathbf{C})$, and then sample $\mathbf{X}_{t-\Delta t}$ from $q(\mathbf{X}_{t-\Delta t}|\mathbf{\hat X}_{0}, \mathbf{X}_{t})$, which involves remask $\lfloor N \cdot \sigma(t-\Delta t)  \rfloor$ tokens in $\mathbf{\hat X}_{0}$ with the lowest confidence score, where we define the confidence score of $\hat{x}_i$ in $\mathbf{\hat X}_{0}$ to  $p_{\theta}(\hat{x}_i|\mathbf{X}_{t}, \mathbf{X}^{p}, \mathbf{C})$ if $m_{t,i} = 1$, otherwise, we set confidence score of $x_i$ to $1$, which means that tokens already unmasked in $\mathcal{X}_{t}$ will not be remasked.

\textbf{Classifier-free Guidance.}
Moreover, we adapt the classifier-free guidance technique \cite{nichol2021glide,ho2022classifier}. Specifically, in training, we do not use the prompt with a probability of $p_{\text{cfg}}=0.15$. In inference, we extrapolate the model output towards the conditional generation guided by the prompt $g_{\text{cond}} = g(\mathbf{X}|\mathbf{X}^{p})$ and away from the unconditional generation $g_{\text{uncond}}=g(\mathbf{X})$, i.e., $g_{\text{cfg}} = g_{\text{cond}} + \alpha \cdot (g_{\text{cond}}-g_{\text{uncond}})$, with a guidance scale $\alpha$ selected based on experimental results.
We then rescale it through $g_{\text{final}} = \text{std}(g_{\text{cond}}) \times g_{\text{cfg}} / \text{std}(g_{\text{cfg}})$, following \cite{lin2024common}.

\subsection{Connections to the NaturalSpeech Series}
\myname{} is an advanced TTS system of the NaturalSpeech series. Compared with the previous versions NaturalSpeech~\cite{tan2024naturalspeech} and NaturalSpeech 2~\cite{shen2023naturalspeech}, \myname{} has the following connections and distinctions: 
\begin{itemize} [leftmargin=*]
    \item \textit{Goal}. The NaturalSpeech series aims to generate natural speech with high quality and diversity. We approach this goal in several stages: 1) Achieving high-quality speech synthesis in single-speaker scenarios. To this end, NaturalSpeech~\cite{tan2024naturalspeech} generates speech with quality on par with human recordings and only tackles single-speaker recording-studio datasets (e.g., LJSpeech). 2) Achieving high-quality and diverse speech synthesis on multi-style, multi-speaker, and multi-lingual scenarios. NaturalSpeech 2~\cite{shen2023naturalspeech} firstly focuses on speech diversity by exploring the zero-shot synthesis ability based on large-scale, multi-speaker, and in-the-wild datasets. Furthermore, \myname{} further achieves human-level naturalness on the multi-speaker dataset (e.g., LibriSpeech).
    
    % Both NaturalSpeech 2~\cite{shen2023naturalspeech} and \myname{} focus on speech diversity by exploring the zero-shot synthesis ability based on large-scale, multi-speaker, and in-the-wild datasets. 
    
    \item \textit{Architecture.} The NaturalSpeech series shares the basic components such as encoder/decoder for waveform reconstruction and duration prediction for non-autoregressive speech generation. Different from NaturalSpeech which utilizes flow-based generative models and NaturalSpeech 2 which leverages latent diffusion models, \myname{} proposes the concept of factorized diffusion models to generate each factorized speech attribute in a divide-and-conquer way.
    
    \item \textit{Speech Representations.} Due to the complexity of speech waveform, the NaturalSpeech series uses an encoder/decoder to obtain speech latent for high-quality speech synthesis. NaturalSpeech utilizes naive VAE-based continuous representations, NaturalSpeech 2 leverages the continuous representations from the neural audio codec with residual vector quantizers, while NaturalSpeech 3 proposes a novel FACodec to convert complex speech signal into disentangled subspaces (i.e., prosody, content, acoustic details, and timbre) and reduces the speech modeling complexity.
\end{itemize}

\section{Experiments and Results}

\subsection{Experimental Settings}

In this subsection, we introduce the training, inference and evaluation for the Factorized Diffusion Model.  Please refer to Appendix \ref{appendix: fdm_config} for model configuration.

\textbf{Implementation Details.}
We use Librilight~\citep{kahn2020libri}, which contains $60$K hours of $16$KHz unlabeled speech data and around 7000 distinct speakers from LibriVox audiobooks, as the training set. In duration diffusion, we further improve the performance by conditioning phoneme-level prosody codes. Specifically, we perform phoneme-level pooling according to duration on the pre-quantized vectors, and then feed these phoneme-level representations into the prosody quantizer in our codec to obtain the phoneme-level prosody codes. We employ an additional discrete diffusion to generate these in inference. We perform $4$ iterations in each diffusion process. We generate duration without classifier-free guidance and generate others with a classifier-free guidance scale of $1.0$. This strategy results in $4 \times 2$ for phoneme-level prosody, $4$ for duration, $4 \times 2$ for each token sequence of prosody, content, and acoustic details, totaling $60$ forward passes due to the double computation with classifier-free guidance. Please refer to Appendix \ref{appendix: codec-implement} for details of the FACodec and Appendix \ref{appendix: fdm_train_infer} for more details of our factorization diffusion model.

\textbf{Evaluation Dataset.} We employ two benchmark datasets: 1) LibriSpeech~\citep{panayotov2015librispeech} test-clean, a widely-used testset for zero-shot TTS task. It contains 40 distinct speakers and 5.4-hour speech. Following \cite{shen2023naturalspeech}, we randomly select one sentence for each speaker for LibriSpeech test-clean benchmark. Specifically, we randomly select $3$-second clips as prompts from the same speaker's speech. 2) RAVDESS~\cite{livingstone2018ryerson}, an emotional TTS dataset featuring 24 professional actors ($12$ female, $12$ male) across $8$ emotions (neutral, calm, happy, sad, angry, fearful, surprise, and disgust) in $2$ emotional intensity (normal and strong). We use strong-intensity samples for RAVDESS benchmark. We adopt this benchmark for prosody evaluation, considering 1) for the same speaker, speech with the same emotion shares similar prosody, while speech with different emotions displays varied prosodies; 2) the benchmark provides speech samples with the same text from the same speaker across eight different emotions.

\textbf{Evaluation Metrics.} Objective Metrics: In the Librispeech test-clean benchmark, we evaluate speaker-similarity (SIM-O and SIM-R), speech quality (UTMOS), and robustness (WER). In specific, 1) for SIM-O and SIM-R, we employ the WavLM-TDCNN\footnote{\url{https://github.com/microsoft/UniSpeech/tree/main/downstreams/speaker_verification}} speaker embedding model to assess speaker similarity between generated samples and the prompt. Results are reported for both similarity to original prompt (SIM-O) and reconstructed prompt (SIM-R); 2) for speech quality, we employ UTMOS~\cite{saeki2022utmos} which is a surrogate objective metric of MOS;
3) for Word Error Rate (WER), we use an ASR model\footnote{\url{https://huggingface.co/facebook/hubert-large-ls960-ft}} to transcribe generated speech. The model is a CTC-based HuBERT pre-trained on Librilight and fine-tuned on the 960 hours training set of LibriSpeech. We also use an advanced ASR model based on transducer~\cite{he2019streaming}\footnote{\url{https://huggingface.co/nvidia/stt_en_conformer_transducer_xlarge}}.
In the RAVDESS benchmark, we evaluate the prosody similarity (MCD and MCD-Acc). In specific, 1) following \cite{al2023nonparallel}, we adopt Mel-Ceptral Distortion (MCD) for prosody evaluation by measuring the differences between generated samples and ground truth samples. We report the results for eight emotions, along with the average result. 2) for MCD-Acc, we evaluate the top-1 emotion accuracy of the generated speech on the RAVDESS benchmark for prosodic similarity measures. Specifically, we adopt a K-Nearest-Neighbors (KNN) model as emotion classifier. We compare MCD distances between the generated speech and the ground-truth speech from the same speaker, across eight different emotions.  Subjective Metrics: We employ comparative mean option score (CMOS) and similarity mean option score (SMOS) in both two benchmarks to evaluate naturalness and similarity, respectively.

\textbf{Evaluation Baselines.}
We compare \myname{} with baselines: 1) VALL-E \cite{wang2023neural}. 2) NaturalSpeech 2 \cite{shen2023naturalspeech}. 3) Voicebox \cite{le2023voicebox}. 4) Mega-TTS 2 \cite{jiang2023mega2}. 5) UniAudio \cite{yang2023uniaudio}. 6) StyleTTS 2 \cite{li2023styletts}. 7) HierSpeech++ \cite{lee2023hierspeech++}. Please refer to Appendix \ref{appendix: fdm_baselines} for details.

\begin{table*}[tb]
\centering
\small
\caption{The evaluation results for \myname{} and the baseline methods on LibriSpeech test-clean. $^\spadesuit$ means the results are obtained from the authors. $^\varheart$ means the results directly obtained from the paper. $^\clubsuit$ means the results are infered from offical checkpoints. $^\vardiamond$ means the reproduced results. Abbreviation: LT (LibriTTS), V (VCTK), LJ (LJSpeech), LL$^\star$ (Librilight Small, Medium), EX (Expresso), MS (MSSS Kor), NI (NIKL Kor). Please refer to Appendix \ref{appendix:more_exp_zeroshottts} for more results on 1) WER inferred by an advanced ASR system, and 2) UTMOS, an automatic metric for MOS.} 
\begin{tabular}{lcccccccc}
\toprule
& Training Data  & Sim-O $\uparrow$    & Sim-R $\uparrow$  & WER$\downarrow$  & CMOS$\uparrow$ & SMOS$\uparrow$  \\
\midrule
Ground Truth & - & 0.68 & - & 1.94  &+0.08 & 3.85 \\
\midrule
VALL-E $^\varheart$  & Librilight & - & 0.58 & 5.90 & - & -  \\ 
VALL-E $^\vardiamond$  & Librilight & 0.47 & 0.51 & 6.11   & -0.60 & 3.46  \\ 
NaturalSpeech 2$^\spadesuit$ & Librilight & 0.55 & 0.62 & 1.94 & -0.18 &3.65\\
Voicebox$^\spadesuit$ & Self-Collected (60kh) & 0.64 & 0.67 & 2.03  & -0.23 &3.69\\
Voicebox$^\vardiamond$ & Librilight & 0.48 & 0.50 & 2.14 & -0.32 &3.52\\
Mega-TTS 2$^\spadesuit$ & Librilight & 0.53 & - & 2.32  & -0.20 &3.63 \\
UniAudio$^\spadesuit$ & Mixed (165kh) & 0.57 & 0.68& 2.49   & -0.25  & 3.71 \\
StyleTTS 2$^\clubsuit$ & LT + V + LJ &0.38 & -& 2.49  & -0.21 & 3.07\\
HierSpeech++$^\clubsuit$ & LT +  LL$^\star$ + EX + MS + NI & 0.51 & -& 6.33& -0.41 &3.50\\  
\midrule
\myname{} & Librilight & \textbf{0.67} & \textbf{0.76} & \textbf{1.81} & ~\textbf{0.00} &\textbf{4.01}\\
\bottomrule
\end{tabular}
\label{table:librispeech_eval}
\end{table*}

\subsection{Experimental Results on Zero-shot TTS}
In this subsection, we compare \myname{} with baselines in terms of: 1) generation quality in Section \ref{sec: qualtiy_eval}; 2) generation similarity in Section \ref{sec: sim_eval}; 3) robustness in Section \ref{sec: robust_eval}. Specifically, for generation similarity, we evaluate in two aspects: 1) speaker similarity; 2) prosody similarity. Please refer to Appendix \ref{appendix: latency} for latency analysis.

\subsubsection{Generation Quality}
\label{sec: qualtiy_eval}
To evaluate speech quality, we conduct CMOS test, with $12$ native as the judges. We randomly select $20$ utterances from both LibriSpeech test-clean and RAVDESS benchmarks. As shown in Table \ref{table:librispeech_eval}, we find that 1) \myname{} is close to the ground-truth recording ($-0.08$ on Librispeech test-clean, and $-0.17$ on RAVDESS), which demonstrates \myname{} can generate high-quality and natural speech; 2) \myname{} outperforms baselines by a substantial margin, verifying the effectiveness of \myname{} with factorization.

\subsubsection{Generation Similarity}
\label{sec: sim_eval}
\textbf{Speaker Similarity.} 
We evaluate the speech similarity with both objective metrics (Sim-O and Sim-R) and subjective metrics (SMOS), with $12$ natives as the judges. We randomly select $10$ utterances for SMOS test. As shown in Table~\ref{table:librispeech_eval}, we find that 1) \myname{} achieves parity in Sim-O and a $0.16$ increase in SMOS with ground truth, which indicates great speaker similarity achieved by our proposed method; 2) \myname{} outperforms all baselines on both objective and subjective metrics, highlighting the superiority of our method with factorization in terms of speaker similarity. Additionally, we notice certain discrepancy between Sim-O and SMOS. For instance, the SMOS is not as competitive as SIM-O for Voicebox model, likely due to some unnatural prosody.

\begin{table}[tb]
\centering
\caption{The evaluation results for \myname{} and the baseline methods on RAVDESS. $^\spadesuit$ means the results are obtained from the authors. $^\clubsuit$ means the results are inferred from official checkpoints. $^\vardiamond$ means the reproduced results. Abbreviation: Avg (average MCD), Acc (MCD-Acc).} 

\begin{tabular}{lcccc}
\toprule
         & Avg$\downarrow$  & Acc$\uparrow$ & CMOS$\uparrow$ & SMOS$\uparrow$  \\
\midrule
Ground Truth & 0.00 & 1.00 & +0.17 &4.42 \\
\midrule
VALL-E $^\vardiamond$ & 5.03 & 0.34 & -0.55 & 3.80 \\
NaturalSpeech 2$^\spadesuit$ & 4.56 & 0.25& -0.22 & 4.04 \\
Voicebox$^\vardiamond$ & 4.88 & 0.34 & -0.34 & 3.92 \\
Mega-TTS 2$^\spadesuit$& 4.44 & 0.39& -0.20 & 4.51\\
StyleTTS 2$^\clubsuit$ & 4.50 & 0.40 & -0.25 &3.98 \\
HierSpeech++$^\clubsuit$ &  6.08 & 0.30 & -0.37 & 3.87 \\ 
\midrule
\myname{}   &  \textbf{4.28} & \textbf{0.52} & ~\textbf{0.00} & \textbf{4.72} \\
\bottomrule
\end{tabular}
\label{table:ravdess_eavl}
\end{table}

\textbf{Prosody Similarity.}
We evaluate prosody similarity with both objective metrics (MCD and MCD-Acc) and subjective metrics (SMOS) on the RAVDESS benchmark. We randomly select $10$ utterances for SMOS test. As shown in Table \ref{table:ravdess_eavl}, \myname{} consistently surpasses baselines by a remarkable margin in MCD avg, MCD-Acc, and SMOS. It reveals that \myname{} achieves a significant improvement in terms of prosodic similarity. Please refer to Appendix \ref{appendix：prosody_detail} for the MCD scores across $8$ emotions.

\subsubsection{Robustness}
\label{sec: robust_eval}
We assess the robustness of our zero-shot TTS by measuring the word error rate of generated speech on the LibriSpeech test-clean benchmark. The results in Table \ref{table:librispeech_eval} indicate that 1) \myname{} achieves a better WER than the ground truth, proving the high intelligibility; 2) \myname{} outperforms other baselines by a considerable margin, which demonstrates the superior robustness of \myname{}.

\subsubsection{Human-Level Naturalness on LibriSpeech Testset}
We compare the speech synthesized by \myname{} with human recordings (Ground Truth) in Table \ref{table:librispeech_eval} (more results can be found in Table \ref{table:librispeech_eval_appendix} in Appendix \ref{appendix:more_exp_zeroshottts}). We have the following observations: 1) \myname{} achieves -0.01 Sim-O and +0.16 SMOS compared to human recordings, which demonstrates that our method is on par or better on speaker similarity; 2) \myname{} achieves -0.08 CMOS and +0.16 UTMOS compared with recording, which demonstrates that our method can generate on-par or better voice quality; 3) Our method also achieves close WER with human recordings, which demonstrates the robustness of \myname{}. Therefore, we can conclude that for the first time, \myname{} has achieved human-level quality and naturalness on the multi-speaker LibriSpeech test set in a zero-shot way. It is another great milestone after NaturalSpeech 1~\cite{tan2024naturalspeech} has achieved human-level quality on the single-speaker LJSpeech dataset.

\subsection{Ablation Study and Method Analyses}

\subsubsection{Ablation Study}
In this subsection, we conduct ablation studies to verify the effectiveness of 1) factorization; 2) classier-free guidance; 3) prosody representation. We also conduct ablation study to compare our duration diffusion model with traditional duration predictor in Appendix \ref{appendix: dur_ablation}. 

\textbf{Factorization.}
To verify the proposed factorization method, we ablate it by removing factorization in both codec and factorized diffusion model. Specifically, we 1) use the discrete tokens from SoundStream, a neural codec which does not consider factorization, and 2) do not consider factorization in generation. As shown in Table \ref{table:ablation_f_cfg}, we could find a significant performance degradation without the factorization, a drop of $0.12$ in Sim-O, $0.15$ in Sim-R, $0.68$ in WER, $0.25$ in CMOS and 0.42 in SMOS. This indicates the proposed factorized method can consistently improve the performance in terms of speaker similarity, robustness, and quality. 

\begin{table}[tb]
\centering
\caption{The ablation study of factorization and classifier-free guidance (cfg) on LibriSpeech test-clean. } 
% \scalebox{0.9}{
\begin{tabular}{lccccc}
\toprule
 & Sim-O / Sim-R $\uparrow$   & WER$\downarrow$ & CMOS$\uparrow$ & SMOS$\uparrow$ \\
\midrule
\myname{} & \textbf{0.67 / 0.76} & \textbf{1.81} & \textbf{0.00}& \textbf{4.01}  \\ 
\midrule
 - factorization  & 0.55 / 0.61  & 2.49 & -0.25  & 3.59 \\
 - cfg  & 0.64 / 0.72 & \textbf{1.81} & -0.06 & 3.80\\
\bottomrule
\end{tabular}
\label{table:ablation_f_cfg} 
\end{table}

\textbf{Classier-Free Guidance.}
We conduct an ablation study by dropping the classifier-free guidance in inference to validate its effectiveness. We double the iterations to ensure the same $60$ forward passes for fair comparison. Table \ref{table:ablation_f_cfg} illustrates a significant degradation without classifier-free guidance, a decrease of $0.03$ in Sim-O, $0.04$ in Sim-R, $0.06$ in CMOS and $0.21$ in SMOS, proving that classifier-free guidance can greatly help the speaker similarity and quality.

\begin{table}[tb]
\centering
\caption{The ablation study of prosody representation on RAVDESS. Denote ``Mel 20 Bins'' using the first 20 bins in the mel-spectrogram as the prosody representation.} 
% \small
\begin{tabular}{lccc}
\toprule
         &MCD Avg$\downarrow$  & MCD-Acc$\uparrow$   \\
\midrule
\myname{}   &  \textbf{4.28} & \textbf{0.52}  \\
Mel 20 Bins & 4.34 & 0.46\\
\bottomrule
\end{tabular}
\label{table: ablation_prosody_rep}
\end{table}

\textbf{Prosody Representation.} We compare different prosody representations on zero-shot TTS task. In specific, we select handcrafted prosody features (e.g., the first 20 bins of mel-spectrogram~\cite{jiang2023mega, oh2023diffprosody, ren2022prosospeech}) as the baseline. We drop the prosody FVQ module and directly quantize the first 20 bins of the mel-spectrogram, without the normalized F0 loss. Table \ref{table: ablation_prosody_rep} shows that using ``Mel 20 Bins'' as prosody representation demonstrates inferiority in terms of prosody similarity compared to the prosody representations learned from codec (4.34 vs 4.28 in average MCD, 0.46 vs 0.52 in MCD-Acc).

\begin{table*}[h!]
\footnotesize
	\centering
 \small
    \caption{The reconstruction quality evaluation of codecs. $^\clubsuit$ means results are infered from offical checkpoints. $^\bigstar$ means the reproduced checkpoint. $^\vardiamond$ means the reproduced model following the original paper's implementation and experimental setup. All models use a codebook size of $1024$. \textbf{Bold} for the best result and \underline{underline} for the second-best result. Abbreviation: H (Hop Size), N (Codebook Number).}
	\begin{tabular}{l c c c c c c c c}
	\toprule
       Models & Sampling Rate & H & N & Bandwidth & PESQ $\uparrow$ & STOI $\uparrow$ & MSTFT $\downarrow$ & MCD $\downarrow$ \\
     \midrule
     EnCodec$^\clubsuit$ & 24kHz & 320 & 8 & 6.0 kbps & 3.28 & 0.94 & 0.99 & 2.70 \\
     HiFi-Codec$^\clubsuit$ & 16kHz & 320 & 4 & 2.0 kbps & 3.17 & 0.93 & 0.98 & 3.05 \\
     DAC$^\clubsuit$ & 16kHz  & 320 & 9 & 4.5 kbps & \textbf{3.52} & \textbf{0.95} & 0.97 & \underline{2.65}\\
    SoundStream$^\vardiamond$ & 16kHz & 200 & 6 & 4.8 kbps & 3.03 & 0.90 & 1.07 & 3.38 \\
    \midrule
    FACodec & 16kHz & 200 & 6 & 4.8 kbps & \underline{3.47} & \textbf{0.95} & \underline{0.93} & \textbf{2.59} \\
    \bottomrule
    \end{tabular}
     \label{tabel:codec_quality_eval_small}
\end{table*}

\subsubsection{Method Analyses}
In this subsection, we first discuss the extensibility of our factorization. We then introduce the application of speech attributes manipulation in a zero-shot way.

\begin{table}[tb]
\centering
\caption{The comparison between autoregressive approach with (VALL-E + FACodec) and without (VALL-E) our proposed factorization on LibriSpeech test-clean. $^\vardiamond$ means the reproduced results. Abbreviation: Sim-O/R (Sim-O / Sim-R).} 

\begin{tabular}{lcccc}
\toprule
 & Sim-O / R $\uparrow$  & WER$\downarrow$ & CMOS$\uparrow$ & SMOS$\uparrow$ \\
\midrule
VALL-E + FACodec  & \textbf{0.57} / \textbf{0.65}  & \textbf{5.60}& \textbf{+0.24} &\textbf{3.61}  \\ 
VALL-E$^\vardiamond$ & 0.47 / 0.51 & 6.11 & 0.00 & 3.46  \\ 
\bottomrule
\end{tabular}
\label{table:extensity}
\end{table}
\textbf{Extensibility.}
\myname{} utilizes a non-autoregressive model for discrete token generation with factorization design. 
To validate the extensibility of our proposed factorization method, we further explore the autoregressive generative model for discrete token generation under our factorization framework. We utilize VALL-E for verification. We first employ an autoregressive language model to generate prosody codes, followed by a non-autoregressive model to generate the remaining content and acoustic details codes. This approach maintains a consistent order of attribute generation, allowing for a fair comparison. We name it VALL-E + FACodec. As shown in Table \ref{table:extensity}, VALL-E + FACodec consistently outperforms VALL-E by a considerable margin in all objective and subjective metrics, demonstrating the factorization design can enhance VALL-E in speech similarity, quality and generation robustness. It further shows our factorization paradigm is not limited in the proposed factorization diffusion model and has a large potential in other generative models. We leave it for future work.

\textbf{Speech Attribute Manipulation.}
As discussed in Section \ref{sec: method-gen}, our factorized diffusion model enables attribute manipulation by selecting different attributes prompts from different speech. We mainly focus on manipulating duration, prosody, and timbre, since the content codes are dictated by the text in TTS, and the acoustic details do not carry semantic information. Leveraging the strong in-context capability of \myname{}, the generated speech effectively mirrors the corresponding speech attributes. For instance, 1) we can utilize the timbre prompt from a different speech to control the timbre while keeping other attributes unchanged; 2) despite the correlation between duration and prosody, we can still solely adjust duration prompt to regulate the speed; 3) moreover, we can combine different speech attributes from disparate samples as desired. This allow us to mimic the timbre while using different prosody and speech speed. Samples are available on our demo page\footnote{\url{https://speechresearch.github.io/naturalspeech3}}.

\subsubsection{Experimental Results on FACodec}
We compare the proposed FACodec in terms of the reconstruction quality with strong baselines, such as EnCodec~\cite{defossez2022high}, HiFi-Codec~\cite{yang2023hifi}, Descript-Audio-Codec (DAC)~\cite{kumar2023high}, and our reproduced SoundStream~\cite{zeghidour2021soundstream}. Table~\ref{tabel:codec_quality_eval_small} shows that our codec significantly surpasses SoundStream in the same bandwidth setting ($0.44$ in PESQ, $0.05 $ in STOI, $0.14$ in MSTFT and $0.79$ in MCD, respectively). Check more details in Appendix \ref{appendix: codec-recon}. Compared with other baselines, FACodec also get comparable performance. Additionally, since our codec decouples timbre information, it can enable zero-shot voice conversion easily, we provide the details and experiment results in Appendix \ref{appendix: codec_zs_vc}. Appendix \ref{appendix: codec-ablation} shows some ablation studies about our FACodec.

\subsection{Effectiveness of Data and Model Scaling} 

In this section, we study the effectiveness of data and model scaling on the proposed factorized diffusion model. We evaluate the zero-shot TTS performance in terms of speaker similarity (Sim-O) and robustness (WER) on an internal test set consisting of $30$ audio clips.

\textbf{Data Scaling.}  With a fixed model size of 500M parameters, we train NaturalSpeech 3, including both FACodec and the factorized diffusion model, on three datasets: 1) a 1K-hour subset randomly drawn from the Librilight dataset, 2) a 60K-hour Librilight dataset, and 3) an internal dataset with 200K hours of speech. In Table \ref{table:scaling_training_hours}, we observe that: 1) even with a mere 1K hours of speech data, our model attains a Sim-O score of $0.64$ and a WER of $3.94$. It shows that with the speech factorization, \myname{} can generate the speech effectively.
2) As we scale up training data from 1K hours to 60K hours, and then to 200K hours, \myname{} displays continuously enhanced performance, with an improvement of $0.08$ and $0.09$ in terms of Sim-O, and $0.91$ and $1.83$ in terms of WER, respectively, thus confirming the benefits of data scaling. 

\textbf{Model Scaling.} We scale up the model size of the factorized diffusion model from 500M to 1B parameters with the internal 200K hours dataset. Specifically, we double the number of transformer layers from $12$ to $24$. The results in Table \ref{table:scaling_model_sizes} show a boost in both speaker similarity ($0.05$ in Sim-O) and robustness ($0.40$ in WER), validating the effectiveness of model scaling. In the future, we will scale up the model size even larger to achieve better results.

\section{Conclusion}
In this paper, we develop a TTS system that consists of 1) a novel neural speech codec with factorized vector quantization (i.e., FACodec) to decompose speech waveform into distinct subspaces of content, prosody, acoustic details and timbre and 2) novel factorized diffusion model to synthesize speech by generating attributes in subspaces with discrete diffusion. \myname{} outperforms the state-of-the-art TTS system on speech quality, similarity, prosody, and intelligibility. We also show that \myname{} can enable speech attribute manipulation, by customizing speech attribute prompts.
Furthermore, we demonstrate that \myname{} achieves human-level performance on the multi-speaker LibriSpeech dataset for the first time and better performance by scaling to 1B parameters and 200K hours of training data. We list the limitations and future works in Appendix \ref{app:limitation_future_works}.

\begin{table}[t]
    \centering
        \caption{The performance of \myname{} on an internal test set, with 500M model size and different hours of training data.}
    \begin{tabular}{ccc}
    \toprule
         & Sim-O$\uparrow$ & WER$\downarrow$  \\
    \midrule
    1K  &  0.64 & 3.94 \\
    60K  & 0.72 & 3.03\\
    200K & \textbf{0.73} & \textbf{2.11}\\
    \bottomrule
    \end{tabular}
    \label{table:scaling_training_hours}
\end{table}

\begin{table}[t]
    \centering
        \caption{The performance of \myname{} on an internal test set, with 200K hours of training data and different model sizes.}
    \begin{tabular}{cccc}
    \toprule
         & Sim-O$\uparrow$ & WER$\downarrow$  \\
    \midrule
    500M  & 0.73 & 2.11 \\
    1B & \textbf{0.78} &\textbf{1.71}\\
    \bottomrule
    \end{tabular}
    \label{table:scaling_model_sizes}
\end{table}

\section{Boarder Impact}
Since our model could synthesize speech with great speaker similarity, it may carry potential risks in misuse of the model, such as spoofing voice identification or impersonating a specific speaker. We conducted the experiments under the assumption that the user agree to be the target speaker in speech synthesis. To prevent misuse, it is crucial to develop a robust synthesized speech detection model and establish a system for individuals to report any suspected misuse.

\bibliographystyle{unsrt}
\bibliography{ref}

%%%%%%%%%%%%%%%%%%%%%%%%%%%%%%%%%%%%%%%%%%%%%%%%%%%%%%%%%%%%%%%%%%%%%%%%%%%%%%%
%%%%%%%%%%%%%%%%%%%%%%%%%%%%%%%%%%%%%%%%%%%%%%%%%%%%%%%%%%%%%%%%%%%%%%%%%%%%%%%
% APPENDIX
%%%%%%%%%%%%%%%%%%%%%%%%%%%%%%%%%%%%%%%%%%%%%%%%%%%%%%%%%%%%%%%%%%%%%%%%%%%%%%%
%%%%%%%%%%%%%%%%%%%%%%%%%%%%%%%%%%%%%%%%%%%%%%%%%%%%%%%%%%%%%%%%%%%%%%%%%%%%%%%
\newpage
\appendix
\onecolumn

\section{Details of Factorization Diffusion Model}

\subsection{Model Configuration} 
\label{appendix: fdm_config}
The phoneme encoder uses a similar architecture as \cite{shen2023naturalspeech} and comprises a $6$-layer Transformer with $8$ attention heads, $512$ embedding dimensions, filter size $2048$ and kernel size $9$ for 1D convolution, and a dropout of $0.1$. In prosody, content and acoustic details diffusion, we adopt a shared $12$-layer Transformer, with $8$ attention heads, $1024$ embedding dimensions, filter size $2048$ and kernel size $3$ for 1D convolution, and a dropout of $0.1$. We additionally use conditional layer normalization in each Transformer block to support diffusion time input. In phoneme-level prosody and duration diffusion, we adopt a $6$-layer Transformer with $8$ attention heads, $1024$ embedding dimensions, filter size $2048$ and kernel size $3$ for 1D convolution, and a dropout of $0.1$. We also use conditional layer normalization in the model to support diffusion time input.

\subsection{Training and Inference Details}
\label{appendix: fdm_train_infer}
We use Librilight~\citep{kahn2020libri}, which contains $60$K hours of $16$KHz unlabeled speech data and around 7000 distinct speakers from LibriVox audiobooks, as the training set. We transcribe using an internal ASR system, convert transcriptions to phonemes via grapheme-to-phoneme conversion~\cite{sun2019token}, and obtain duration with an internal alignment tool.  We use $8$ A100 80GB GPUs with a batch size of $10$K frames of latent vectors per GPU for $1$M steps. We use the AdamW optimizer with a learning rate of $1e-4$, $\beta_1 = 0.9$, and $\beta_2 = 0.98$, $5$K warmup steps following the inverse square root learning schedule. 

During inference, we perform $4$ iterations in each diffusion process, including phoneme-level prosody, duration, prosody, content and acoustic details diffusion. We generate duration without classifier-free guidance, and generate others with a classifier-free guidance scale of $1.0$. This strategy results a $4 \times 2$ for phoneme-level prosody, $4$ for duration, $4 \times 2$ for each token sequence of prosody, content and acoustic details, totaling $60$ forward passes due to the double computation with classifier-free guidance. We use a top-k of $20$, with sampling temperature annealing from $1.5$ to $0$. Following \cite{chang2022maskgit}, Gumbel noises are added to token confidences when determining which positions to re-mask in $q(\mathbf{X}_{t-\Delta t}|\mathbf{\hat X}_{0}, \mathbf{X}_{t})$, mentioned in Section \ref{sec:method-diffusion-formulation}.

\subsection{Evaluation Baselines}
\label{appendix: fdm_baselines}
We compare \myname{} with following strong zero-shot TTS baselines:
\begin{itemize}
    \item VALL-E \cite{wang2023neural}. It use an autoregressive and an additional non-autoregressive model for discrete token generation. We report the scores directly obtained from the paper.  We additionally reproduce it using discrete tokens from SoundStream on Librilight. 
    \item NaturalSpeech 2 \cite{shen2023naturalspeech}. It use a non-autoregressive model for continuous vectors generation. We obtain samples through communication with the authors.
    \item Voicebox \cite{le2023voicebox}.  It use a non-autoregressive model for continuous vectors generation. We obtain samples through communication with the authors.  We additionally reproduce it using mel-spectrogram on Librilight.
    \item Mega-TTS 2 \cite{jiang2023mega2}. It use a non-autoregressive model for continuous vectors generation. We obtain samples through communication with the authors.
    \item UniAudio \cite{yang2023uniaudio}. It use an autoregressive model for discrete token generation. We obtain samples through communication with the authors.
    \item StyleTTS 2 \cite{li2023styletts}. It use a non-autoregressive model for continuous vectors generation. We use official code and checkpoint\footnote{\url{https://github.com/yl4579/StyleTTS2}}.
    \item HierSpeech++ \cite{lee2023hierspeech++}. It use a non-autoregressive model for continuous vectors generation. We use official code and checkpoint\footnote{\url{https://github.com/sh-lee-prml/HierSpeechpp}}. We do not use its super resolution model for fair comparison. 
\end{itemize}

\subsection{More Experimental Results on Zero-shot TTS}
\label{appendix:more_exp_zeroshottts}

In this section, we report more evaluation results for \myname{} and other baselines on: 1) WER, inferred by an advanced ASR system\footnote{\url{https://huggingface.co/nvidia/stt_en_conformer_transducer_xlarge}}; 2) UTMOS~\cite{saeki2022utmos}, which is a surrogate objective metric of MOS. The results are shown in Table \ref{table:librispeech_eval_appendix}.

\begin{table*}[tb]
\centering
\small
\caption{The evaluation results for \myname{} and the baseline methods on LibriSpeech test-clean. $^\spadesuit$ means the results are obtained from the authors. $^\varheart$ means the results directly obtained from the paper. $^\clubsuit$ means the results are inferred from offical checkpoints. $^\vardiamond$ means the reproduced results. WER$^\star$ means the word error rate calculated by an advanced ASR system mentioned in ~\ref{appendix:more_exp_zeroshottts}.} 
\begin{tabular}{lccccccc}
\toprule
 & Sim-O $\uparrow$    & Sim-R $\uparrow$  & WER$\downarrow$&WER$^\star$ $\downarrow$ & UTMOS $\uparrow$  & CMOS$\uparrow$ & SMOS$\uparrow$  \\
\midrule
Ground Truth  & 0.68 & - & 1.94 & 0.68 & 4.14 &+0.08 & 3.85 \\
\midrule
VALL-E $^\varheart$   & - & 0.58 & 5.90 &-&-& - & -  \\ 
VALL-E $^\vardiamond$ & 0.47 & 0.51 & 6.11  & 4.87& 3.68 & -0.60 & 3.46  \\ 
NaturalSpeech 2$^\spadesuit$ & 0.55 & 0.62 & 1.94 & 1.24 & 3.88 & -0.18 &3.65\\
Voicebox$^\spadesuit$  & 0.64 & 0.67 & 2.03 & 1.81& 3.82 & -0.23 &3.69\\
Voicebox$^\vardiamond$ & 0.48 & 0.50 & 2.14 & 1.24 & 3.73 & -0.32 &3.52\\
Mega-TTS 2$^\spadesuit$  & 0.53 & - & 2.32 & 2.17& 4.02 & -0.20 &3.63 \\
UniAudio$^\spadesuit$ & 0.57 & 0.68& 2.49  & 1.81& 3.79 & -0.25  & 3.71 \\
StyleTTS 2$^\clubsuit$ &0.38 & -& 2.49 & 1.58 & 3.94 & -0.21 & 3.07\\
HierSpeech++$^\clubsuit$  & 0.51 & -& 6.33& 4.97& 3.80 & -0.41 &3.50\\  
\midrule
\myname{} &  \textbf{0.67} & \textbf{0.76} & \textbf{1.81} & \textbf{1.13}& \textbf{4.30}& ~\textbf{0.00} &\textbf{4.01}\\
\bottomrule
\end{tabular}
\label{table:librispeech_eval_appendix}
\end{table*}

\subsection{Latency Analysis}
\label{appendix: latency}

In this subsection, we compare the inference latency of \myname{} with an autoregressive method (VALL-E) and a non-autoregressive method (NaturalSpeech 2). We also investigate the effect of reducing the number of iterations in each diffusion from 4 to 1, resulting in a total of 15 forward passes. We call this variant \myname{} one-step.  We evaluate the performance on Librispeech test-clean in terms of speaker similarity (Sim-O/Sim-R) and quality (UTMOS \cite{saeki2022utmos} \footnote{\url{https://github.com/tarepan/SpeechMOS}}, a surrogate objective metric of CMOS). The latency tests are conducted on a server with E5-2690 Intel Xeon CPU, 512GB memory, and one NVIDIA V100 GPU. The results are shown in Table \ref{table: latency}. From the results, we have several observations. 1) \myname{} achieves a $15.27\times$ speedup over VALL-E and $1.24\times$ speedup over NaturalSpeech 2, while consistently surpasses these baselines on all metrics. This demonstrate \myname{} is both effective and efficient. 2) when using fewer diffusion steps, \myname{} can still maintain robust performance ($-0.01$ in Sim-O, $-0.01$ in Sim-R, and $-0.29$ in UTMOS) with a $4.41\times$ faster speed, proving the robustness of diffusion steps.

\begin{table}[tb]
\centering
\caption{The latency study on LibriSpeech test-clean. \myname{} one-step denotes using only $1$ iteration in each diffusion process instead of original $4$. Abbreviation: NFE (number of function evaluation).} 

\begin{tabular}{lccccc}
\toprule
Models  & NFE & RTF $\downarrow$ & Sim-O $\uparrow$ & Sim-R $\uparrow$   & UTMOS $\uparrow$ \\
\midrule
NaturalSpeech 2 & 150 & 0.366&  0.55 & 0.62 &  3.87\\
VALL-E & - & 4.520 & 0.47 & 0.51 & 3.67 \\
\midrule
\myname{} & 60 &  0.296 & \textbf{0.67} & \textbf{0.76} & \textbf{4.30} \\
\myname{} one-step & 15 & \textbf{0.067} &  0.66 & 0.75 & 4.01 \\

\bottomrule
\end{tabular}
\label{table: latency}
\end{table}

\subsection{Ablation Study on Duration Diffusion Model}

\label{appendix: dur_ablation}
In this subsection, we conduct an ablation study to compare our duration discrete diffusion model with the traditional duration predictor, which regresses the duration in logarithmic domain. The ablation study focus on 1) Generation: multi-step generation vs. one-step generation. 2) Objective: classification-based cross-entropy loss vs. regression-based L2 loss. 3) Conditioning: with vs. without phoneme-level prosody conditioning. 4) Prompting: with vs. without duration prompting.  We evaluate them on Librispeech test-clean in terms of speaker similarity (Sim-O/Sim-R), robustness (WER) and qualtiy (UTMOS).  As shown in Table \ref{table:dur_ablation}, we can find that 1) without multi-step generation, there’s a significant drop in performance (-0.05 in Sim-O, -0.03 in Sim-R, and -0.12 in UTMOS).  2) replacing cross-entropy loss with l2 loss affects the performance, causing a decrease of -0.05 in Sim-O, -0.04 in Sim-R, 0.44 in WER and -0.17 in UTMOS. 3) dropping phoneme-level prosody conditioning will affect both speaker similarity (-0.05 in Sim-O and -0.04 in Sim-R), robustness (0.55 in WER) and quality (-0.19 in UTMOS) 4) the duration prompting mechanism is crucial for speaker similarity, robustness and quality, with changes of -0.06 in Sim-O, -0.05 in Sim-R, 0.89 in WER and -0.22 in UTMOS. These results confirm that each design aspect of our duration predictor contributes to performance improvement.

\begin{table*}[tb]
\centering
\caption{The ablation results of the design of the duration predictor on LibriSpeech test-clean.} 

\begin{tabular}{lcccc}
\toprule
         & Sim-O $\uparrow$    & Sim-R $\uparrow$  & WER$\downarrow$  & UTMOS$\uparrow $    \\
\midrule
\myname{} & \textbf{0.67} & \textbf{0.76} & \textbf{1.94} & \textbf{4.30}\\
\midrule
Generation ablation & 0.62 & 0.73 & \textbf{1.94} & 4.18  \\
Objective ablation& 0.62 & 0.72 & 2.38 &4.13\\
Conditioning ablation& 0.62 & 0.72 & 2.49 &4.11\\
Prompting ablation& 0.61 & 0.71& 2.83& 4.08 \\
\bottomrule
\end{tabular}
\label{table:dur_ablation}
\end{table*}

\subsection{Details of Prosody Similarity Evaluation }
\label{appendix：prosody_detail}
 In Table \ref{table: prosody_detail}, we present MCD on $8$ different emotions, comparing \myname{} with the baseline methods on the RAVDESS benchmark. \myname{} demonstrates robust performance across $8$  emotions, verifying the effectiveness and robustness in terms of prosody similarity.

\begin{table*}[tb]
\centering
\caption{The MCD scores on $8$ different emotions of \myname{} and the baseline methods on RAVDESS. $^\spadesuit$ means the results are obtained from the authors. $^\clubsuit$ means the results are inferred from official checkpoints. $^\vardiamond$ means the reproduced results. We use \textbf{bold} to indicate the best result and \underline{underline} to indicate the second-best result.} 

\begin{tabular}{lccccccccc}
\toprule
         & \multicolumn{8}{c}{MCD$\downarrow$}    \\
         & neutral  &    calm  &   happy  &     sad  &   angry &  fearful &  disgust & surprised   \\
\midrule
Ground Truth & 0.00 & 0.00 &  0.00 & 0.00 &  0.00 &  0.00 & 0.00 &  0.00  \\
\midrule
% YourTTS  &     &              \\
VALL-E $^\vardiamond$  & 3.97 & 4.75 & 4.83 & 5.51 & 5.19 & 5.29 & 5.45 & 5.29  \\ 
Voicebox$^\vardiamond$ & 3.93 & 4.90 & 4.96 & 4.93 & 5.01 & 5.03 & 5.34 & 4.89  \\
% Voicebox$^\spadesuit$  & \\
NaturalSpeech 2$^\spadesuit$ & \textbf{2.77} & \textbf{3.51} &  4.85 & 4.88 & 5.42 & 5.23 & 5.31 & 4.52\\
Mega-TTS 2$^\spadesuit$ &  3.28 & 4.39 & 4.44 & 4.67 & \textbf{4.21}& 5.00 & 5.42 & \textbf{4.14} \\
StyleTTS 2$^\clubsuit$  & 3.41 & 4.38 & \underline{4.40} & \underline{4.64} & 4.80 & \underline{4.69} & \underline{5.10} & 4.57  \\
HierSpeech++$^\clubsuit$  & 5.54 & 6.55 & 5.78 & 5.84 & 6.37 & 6.17 & 6.74 & 5.62  \\ 
\midrule
\myname{}  & \underline{3.23} & \underline{4.32} & \textbf{4.26} & \textbf{4.41} & \underline{4.64} & \textbf{4.25} & \textbf{4.80} & \underline{4.45}  \\
\bottomrule
\end{tabular}
\label{table: prosody_detail}
\end{table*}

\section{Details of FACodec}

\subsection{Implementation Details}
\label{appendix: codec-implement}

\textbf{Model Architecture.}
The basic architecture of FACodec encoder and decoder follows \cite{kumar2023high} and employs the SnakeBeta activation function~\cite{lee2022bigvgan}. The timbre extractor consists of several conformer~\cite{gulati2020conformer} blocks. We use $N_{q_c} = 2, N_{q_p} = 1, N_{q_d} = 3$ as the number of quantizers for each of the three FVQ $\mathcal{Q}^c, \mathcal{Q}^p, \mathcal{Q}^d$, the codebook size for all the quantizers is 1024.

\textbf{Loss Functions.} 
We utilize the multi-scale mel-reconstruction loss $\mathcal{L}_{\text{rec}}$ as detailed in \cite{kumar2023high}. For the adversarial loss $\mathcal{L}_{\text{adv}}$, we employ both the multi-period discriminator (MPD) and the multi-band multi-scale STFT discriminator, as proposed by \cite{kumar2023high}. Additionally, we incorporate the relative feature matching loss $\mathcal{L}_{\text{feat}}$. For codebook learning, we use the codebook loss $\mathcal{L}_{\text{codebook}}$ and the commitment loss $\mathcal{L}_{\text{commit}}$ from VQ-VAE~\cite{van2017neural}. The training loss also includes the phone prediction loss $\mathcal{L}_{\text{ph}}$, the normalized F0 prediction loss $\mathcal{L}_{\text{f0}}$, and the gradient reverse losses of phone prediction $\mathcal{L}_{\text{gr-ph}}$, normalized F0 prediction $\mathcal{L}_{\text{gr-f0}}$, and speaker classification $\mathcal{L}_{\text{gr-spk}}$ for disentanglement learning. The total training loss for the generator can be formulated as:
   $\lambda_{\text{rec}}\mathcal{L}_{\text{rec}} + \lambda_{\text{adv}}\mathcal{L}_{\text{adv}} + \lambda_{\text{feat}}\mathcal{L}_{\text{feat}} + \lambda_{\text{codebook}}\mathcal{L}_{\text{codebook}} 
+ \lambda_{\text{commit}}\mathcal{L}_{\text{commit}} + \lambda_{\text{ph}}\mathcal{L}_{\text{ph}} + \lambda_{\text{f0}}\mathcal{L}_{\text{f0}} + \lambda_{\text{gr-ph}}\mathcal{L}_{\text{gr-ph}} 
 + \lambda_{\text{gr-f0}}\mathcal{L}_{\text{gr-f0}} + \lambda_{\text{gr-spk}}\mathcal{L}_{\text{gr-spk}},$ where $\lambda_{\text{rec}}$, $\lambda_{\text{adv}}$, $\lambda_{\text{feat}}$, $\lambda_{\text{codebook}}$, $\lambda_{\text{commit}}$,$ \lambda_{\text{f0}}$, $\lambda_{\text{ph}}$, $\lambda_{\text{gr-f0}}$, $\lambda_{\text{gr-ph}}$ and $\lambda_{\text{gr-spk}}$ are coefficients for balancing each loss terms. In our paper, we set these coefficients as follows: $\lambda_{\text{rec}}=10.0$, $\lambda_{\text{adv}}=2.0$, $\lambda_{\text{feat}}=2.0$, $\lambda_{\text{codebook}}=1.0$, $\lambda_{\text{commit}}=0.25$, $\lambda_{\text{f0}}=5.0$, $\lambda_{\text{ph}}=5.0$, $\lambda_{\text{gr-f0}}=5.0$, $\lambda_{\text{gr-ph}}=5.0$ and $\lambda_{\text{gr-spk}}=1.0$.

\textbf{Training Details.}
We use Librilight as the training set. We train the codec using 8 NVIDIA TESLA V100 32GB GPUs with a batch size of 32 speech clips of 16000 frames each per GPU for $800$K steps. We use the Adam optimizer with a learning rate of $2e-4$, $\beta_1 = 0.5$, and $\beta_2 = 0.9$.

\subsection{Reconstruction Performance Comparison}
\label{appendix: codec-recon}

We evaluate the reconstruction performance with the following objective metrics: Perceptual Evaluation of Speech Quality (PESQ), Short-Time Objective Intelligibility (STOI), Multi-Resolution STFT Distance (MSTFT), and Mel-Cepstral Distortion (MCD). These metrics collectively measure the difference between the original and the reconstructed samples. We select the following open-source codec models as baselines: EnCodec~\cite{defossez2022high}\footnote{\url{https://github.com/facebookresearch/encodec}}, HiFi-Codec~\cite{yang2023hifi}\footnote{\url{https://github.com/yangdongchao/AcademiCodec}}, and Descript-Audio-Codec (DAC)~\cite{kumar2023high}\footnote{\url{https://github.com/descriptinc/descript-audio-codec}}. We additionally reproduce SoundStream~\cite{zeghidour2021soundstream} following the original paper's implementation and experimental setup. Table ~\ref{codec_quality_eval} shows that 1) FACodec significantly surpasses SoundStream in the same bandwidth setting ($0.44$ in PESQ, $0.05 $ in STOI, $0.14$ in MSTFT and $0.79$ in MCD, respectively). Moreover, FACodec achieves on-par performance with SoundStream even when its bandwidth is doubled ($0.02$ in PESQ, $0.01$ in STOI, $-0.01$ in MSTFT and $0.17$ in MCD, respectively). 2) For a fair comparison, we compare FACodec with other baselines in a similar bandwidth. FACodec achieve comparable or better result on most metrics than these strong baselines, which means that we can still achieve excellent reconstruction speech quality when disentangling speech attributes.

\begin{table*}[h!]
\footnotesize
	\centering
    \small                    
    \caption{The reconstruction quality evaluation of codecs. $^\clubsuit$ means results are infered from offical checkpoints. $^\bigstar$ means the reproduced checkpoint. $^\vardiamond$ means the reproduced model following the original paper's implementation and experimental setup. All models use a codebook size of $1024$. We use \textbf{bold} to indicate the best result and \underline{underline} to indicate the second-best result. Abbreviation: H (Hop Size), N (Codebook Number).}
    % \resizebox{\textwidth}{!}{
	\begin{tabular}{l c c c c c c c c}
	\toprule
       Models & Sampling Rate & H & N & Bandwidth & PESQ $\uparrow$ & STOI $\uparrow$ & MSTFT $\downarrow$ & MCD $\downarrow$ \\
     \midrule
     EnCodec$^\clubsuit$ & 24kHz & 320 & 8 & 6.0 kbps & 3.28 & 0.94 & 0.99 & 2.70 \\
     EnCodec$^\bigstar$ & 16kHz & 320 & 10 & 5.0 kbps & 3.10 & 0.92 & 0.97 & 3.10 \\
     HiFi-Codec$^\clubsuit$ & 16kHz & 320 & 4 & 2.0 kbps & 3.17 & 0.93 & 0.98 & 3.05 \\
     DAC$^\clubsuit$ & 16kHz  & 320 & 9 & 4.5 kbps & \textbf{3.52} & \textbf{0.95} & 0.97 & \underline{2.65}\\
    \midrule
    SoundStream$^\vardiamond$ & 16kHz & 200 & 6 & 4.8 kbps & 3.03 & 0.90 & 1.07 & 3.38 \\
    SoundStream$^\vardiamond$ & 16kHz & 200 & 12 & 9.6 kbps & 3.45 & 0.94 & \textbf{0.92} & 2.76 \\
    \midrule
    FACodec & 16kHz & 200 & 6 & 4.8 kbps & \underline{3.47} & \textbf{0.95} & \underline{0.93} & \textbf{2.59} \\
    \bottomrule
    \end{tabular}
    % }
     \label{codec_quality_eval}
\end{table*}

\subsection{Zero-shot Voice Conversion}
\label{appendix: codec_zs_vc}
Voice conversion aims to transform speech from a source speaker into that of a target speaker, preserving content while altering timbre. Zero-shot voice conversion achieves this by utilizing a prompt speech sample from the target speaker to convert the source speaker's speech. FACodec achieves zero-shot voice conversion by extracting the speaker embedding $h_t^{prompt}$ from the prompt speech to replace the speaker embedding $h_t^{source}$ from the source speech, and utilizing content codes $z_c^{source}$, prosody codes $z_p^{source}$, and detail codes $z_d^{source}$ from the source speaker to reconstruct the target speech $\mathcal{D}(z_c^{source}, z_p^{source}, z_d^{source}, h_t^{prompt})$. We compare FACodec with some previous SOTA models: YourTTS~\cite{casanova2022yourtts}, Make-A-Voice (VC)~\cite{huang2023make}, LM-VC~\cite{wang2023lm}, and UniAudio~\cite{yang2023uniaudio}. We use VCTK dataset for comparison. We use Sim-O\footnote{\url{https://huggingface.co/microsoft/wavlm-base-plus-sv}} to compare speaker similarity to baselines and WER to evaluate speech quality. Table~\ref{vc_exp} shows the evaluation results. The experimental results demonstrate that FACodec solely achieves comparable similarity and superior intelligence compared to the state-of-the-art zero-shot VC models, which need additional training on this task. This implies that FACodec achieves superior disentanglement, especially in timbre.

\begin{table*}[h!]
\small
	\centering
    \caption{The zero-shot voice conversion evaluation results for FACodec with previous SOTA methods. We use \textbf{bold} to indicate the best result and \underline{underline} to indicate the second-best result.}
	\begin{tabular}{l c c}
	\toprule
       Models & Sim-O $\uparrow$ & WER $\downarrow$ \\
     \midrule
        Ground Truth & - & 3.25 \\
     \midrule
        YourTTS & 0.72 & 10.1\\
        Make-A-Voice (VC) & 0.68 & 6.20\\
        LM-VC & 0.82 & 4.91\\
        UniAudio & \textbf{0.87} & \underline{4.80}\\
    \midrule
    FACodec & \underline{0.86} & \textbf{3.46}\\
    \bottomrule
    \end{tabular}
     \label{vc_exp}
\end{table*}

\subsection{Ablation Study}
\label{appendix: codec-ablation}

In this subsection, we study 1) the impact of the information bottleneck on the disentanglement of our FACodec; 2) the effect of gradient reversal on the disentanglement of our FACodec; 3) the role of the acoustic details quantizers; 4) the effects of different prosody representations for TTS generation.

\textbf{Information Bottleneck for Disentanglement.}

We investigate the impact of the information bottleneck on speech disentanglement through qualitative analysis. We find that without using information bottleneck (quantize in original dimensional space rather than low dimensional space) can lead to incomplete disentanglement. For example, we conduct zero-shot voice conversion in the same experimental setting using the FACodec without information bottleneck, as mentioned in Appendix~\ref{appendix: codec_zs_vc}. We observe that the timbre of the converted speech is the interpolation between the source and target, indicating its poor timbre disentanglement. Table~\ref{ablation_codec_information} demonstrates that without the information bottleneck, the speaker similarity of zero-shot voice conversion decreases by 0.13.

\begin{table*}[h!]
\small
	\centering
    \caption{Comparison of zero-shot voice conversion evaluation results for FACodec with and without using information bottleneck.}
	\begin{tabular}{l c}
	\toprule
       & Sim-O $\uparrow$ \\
     \midrule
    w. information bottleneck & \textbf{0.86} \\
    w.o. information bottleneck & 0.73 \\
    \bottomrule
    \end{tabular}
     \label{ablation_codec_information}
\end{table*}

\textbf{Gradient Reversal for Disentanglement.}

We investigate the impact of gradient reversal on the disentanglement of the FACodec through qualitative analysis. We observe that not using gradient reversal diminishes the disentangling ability of FACodec.  For instance, removing the content and prosody gradient reversal from the acoustic detail module results in some content and prosody information leaking into the detail acoustic. We can confirm this by solely reconstructing the speech using detail codes and timbre embedding, where partial content and pitch variations can be heard.

\textbf{Role of Acoustic Details Quantizer.}

Although content, prosody, and timbre information already encompass the majority of speech information, Table~\ref{table: codec_detail} demonstrates that employing acoustic details quantizers enhances the speech reconstruction quality of FACodec. We find 1) without using acoustic details quantizers (only utilizing three codebooks), FACodec achieves comparable or better results compared to SoundStream with using three codebooks, which means that content codes, prosody codes, and timbre embedding already contain most of the necessary information for speech reconstruction; 2) adding acoustic details achieves better reconstruction quality, which suggests that acoustic details codes primarily serve to supplement high-frequency details.

\begin{table*}[h!]
\small
	\centering
    \caption{The reconstruction quality comparison between our FACodec with and without using acoustic details quantizers.}
	\begin{tabular}{l c c c c c }
	\toprule
        & Codebook Number & PESQ $\uparrow$ & STOI $\uparrow$ & MSTFT $\downarrow$ & MCD $\downarrow$ \\
     \midrule
    FACodec &  6 &  \textbf{3.47} &  \textbf{0.95} & 
    \textbf{0.93} & \textbf{2.59} \\
    - acoustic details quantizers &  3 &  \underline{3.09} &  \underline{0.92} &  1.08 & \underline{3.12} \\
    \midrule
    SoundStream & 6 & 3.03 & 0.90 & \underline{1.07} & 3.38 \\
    \bottomrule
    \end{tabular}
     \label{table: codec_detail}
\end{table*}

\section{Limitation and Future Works}
\label{app:limitation_future_works}
Despite our proposed TTS system has achieved great progress, we still have the following limitations:

\textbf{Attribute Coverage.} In this work, we propose the factorization design for speech representation and generation, and have achieved significant improvement by factorizing speech into content, prosody, duration, acoustic details and timbre. However, these attributes can not coverage all speech aspects. For example, we can not extract the background sounds, which is a common challenge for speech disentanglement. In the future, we will explore more attributes including: 1. energy, 2. background sounds, and etc.

\textbf{Data Coverage.} Although we have achieved remarkable improvement on zero-shot speech synthesis on speech quality, similarity and robustness, \myname{} is trained on English corpus from LibriVox audiobooks. Thus, it can not coverage real word people's diverse voice and can not support multilingual TTS. In the future, we will address this limitation by collecting more speech data with larger diversity.

\textbf{Neural Speech Codec.} Although our FACodec can factorize speech into attributes and reconstruct with high quality, it still has the following limitations: 1) we need phoneme transcription for content supervision, which limits the scalability; 2) we only verified the disentanglement in zero-shot TTS task. In the future, firstly, we will explore more general methods to achieve better disentanglement, especially without supervision. Secondly, we would like to explore more tasks with the FACodec, such as zero-shot voice conversion and automatic speech recognition.

\end{document}